# A Henon map for the transmission dynamics of COVID-19: The role of asymptomatic transmitters and delayed symptoms

Akshay Pal and Jayanta K Bhattacharjee

School of Physical Sciences

Indian Association for the Cultivation of Science

Jadavpur, Kolkata 700032

## ABSTRACT

We consider the transmission dynamics of COVID-19 which is characterized by two distinct features. One is the existence of asymptomatic carriers which is a hidden variable in the problem. The other is the issue of latency which means that among the symptomatic carriers there could be a fraction whose symptoms develop after a couple of days. Our modelling is restricted to what we call the Phase -1 of the disease. During this phase the disease sets in and the number of infected people starts growing fast ( the number of new cases per day keeps growing on an average ) and then it slows down ( the number of new cases per day starts decreasing ) with the number of new cases decreasing to about one tenth of its peak value or even smaller). We define Phase-1 to be over when the daily cases start rising once again. We write down a Henon-like map to take various effects into account for this first phase. Our principle findings are:

i) The initial growth rate of symptomatic patients is slow if the probability of resulting in an asymptomatic patient as a result of contact between a healthy person and an infected person is high

ii) If the testing and isolating efforts of the government can lead to successful quarantining of the symptomatic patient then the evolution of the disease will be checked very effectively even if the quarantining of asymptomatics is less than successful

iii) By comparing the evolution dynamics of our map with the evolution dynamics of a country like New Zealand we conclude that the ratio of asymptomatic to symptomatic patients will be close to unity

iv) The same procedure for a typical European country like Austria shows that this ratio will be a factor of two or higher

v) Specifically, for India we predict that the Phase 1 will continue to the middle of March 2021 leading to a total number of $(12.5 \pm 0.7) \times 10^6$ symptomatic patients.

vi) The latency effect will cause an increase in the initial growth rate but a slower approach to the final steady state

# I  INTRODUCTION:

In December 2019 a new epidemic started spreading across the world with two unique features. To recall the simplest scenario for an epidemic, consider a community with two kinds of people on a given ( let's say the nth ) day- $x_n$, the fraction of healthy (virus free) but susceptible people and $y_n$ the fraction of infected people. If we ignore the number of deaths and assume that on a given day the number of new infected people will be jointly proportional to the existing number of infected and uninfected people, we get the logistic map [1] describing the spreading of a standard influenza virus. More sophisticated versions of this are the Ross model [2] and the SIR model [3]. With this new virus (COVID-19) there are two dangerous twists.

A) There are two kinds of infected people- one group which does not show any symptoms but can pass it on over the next couple of weeks (i.e. several days). These are the asymptomatic spreaders – the hidden variable in this problem.

B) The other twist is that those who are symptomatic may take a certain number of days (could be as long as three or four) to develop the symptoms. Hence this group will have been infecting people not just on the n-th day but also on n-1, n-2 th day if the latency period is taken to be 3 days. This latency period implies that any map (or differential equation) will have to have an in-built delay.

The standard literature [4-6] of the dynamics of the spreading of infection has been very nicely summarised in Ref 7. Extension of the SIR models to SEIR models have been carried out in Refs [8-10] and the number of deaths have been included in the SEIRD model of Ref [11]. The dynamics is traditionally written in terms of differential equations. The simplest is the logistic equation [1] corresponding to the map described above. The more complicated variety ( the analogue of the Lotka- Volterra population dynamics ) is the Ross model [2] for spreading of malaria and the popular SIR ( susceptible- infected- recovered ) developed by Kermach and McKendrick (3). None of these models include the two vital issues mentioned above. In the SEIR ( susceptible-exposed-infected-recovered ) model [8-10] where a distinction is made between the susceptible population and those that have been exposed to the disease but are not infectious. A significant variant of the SIR model is the stochastic SIR model [12,13] which is an important variation and was used in the early stages of the present disease [14-15]. Variations of the stochastic SIR model have been recently studied to model the dynamics of the disease in China [16] and make predictions about its spread in India [17]. More recently a model consisting of a set of delay differential equations was written down which includes all the new features mentioned above [18]. A simplified version which is a logistic delay equation with the various features of the epidemic included and particularly suitable for taking into account the interventions on the part of the government ( e.g. imposition of lockdown, requirement of social distancing etc ) has been studied in [19,20] . An interesting variant, starting from a different premise, is the work of Cherednik [21] and a discussion of the early mathematical models pertaining to COVID -19 can be found in Ref [22]. A very early analysis of the importance of the asymptomatic transmitters [23] in the South African context was done by Anguelov et al [ 24]. The most detailed models are the agent-based models [30] which in principle can capture every aspect of this problem. However, they require a huge amount of computer time to be reasonably useful.

As far as we can tell, maps have not been used to model this particular dynamic except for the work of Jahedi and Yorke [25] where a one-dimensional non-autonomous logistic map was used to suggest "the best pandemic models are the simplest"! In this work, we will write down a Henon like map for the spreading of this infection with the existence of asymptomatic but infected people taken into account. Our goal is simple. We want to point out that the inclusion of the asymptomatic infected drastically alters the fixed-point structure and clearly indicates the very important role of this hidden variable in the transmission dynamics of this disease problem.

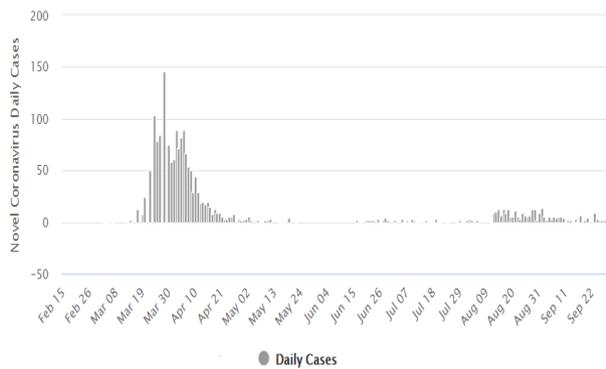

Fig 1a: - (Type 1 country) New Zealand daily cases.

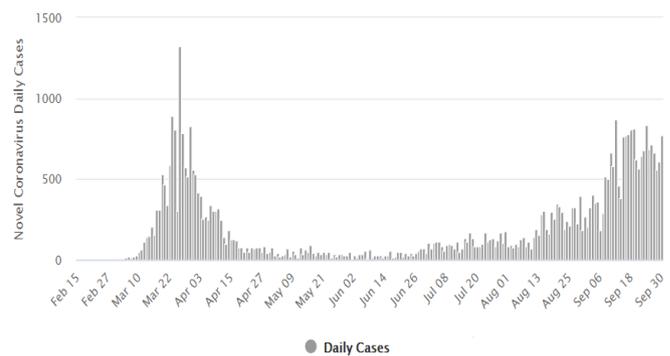

Fig 1b: -(Type 2 country) Austria daily cases.

Looking at the first four months of the spread of this disease ( approximately from early March 2020 to early July 2020 ) we find that for a large number of countries, the total number $N(t)$ of infected people ( time is measured in days ) typically shows two different kinds of behaviour. In both cases there is an initial fast rise. Subsequently 'flattening of the curve" occurs and the rate $dN/dt$ ( physically the number of people infected on a given day) deceases. In some cases ( Type -1 countries in our work) the rate actually goes to zero and stays at zero for a few days and then starts rising again ( Fig 1a). In other cases ( we call them Type -2 countries), $dN/dt$ decreases to a very small value and remains at that small number for a few weeks and then the slope changes to a larger value ( Fig 1b). In both cases the day at which the slope starts increasing again (whether from zero or from a small positive value) marks the end of "Phase -1" ( our terminology) of the spread of this disease. After a brief lull, the number of infected persons has been rising very strongly again and this is in our nomenclature Phase -2. It should be noted that several countries like South Africa, USA, Mexico, Brazil and India have not reached the end of Phase -1 yet. However, countries like South Africa, Brazil and India appear to head for a reasonable completion to Phase-1 in a few months and we can make some predictions about them. Countries like USA have seen a very strong resurgence in the number of cases well before the completion of Phase-1 and is not a part of our modelling. Among the type -1 countries are New Zealand, Switzerland etc and among the Type-2 countries are Austria, the Scandinavian countries, UK etc.

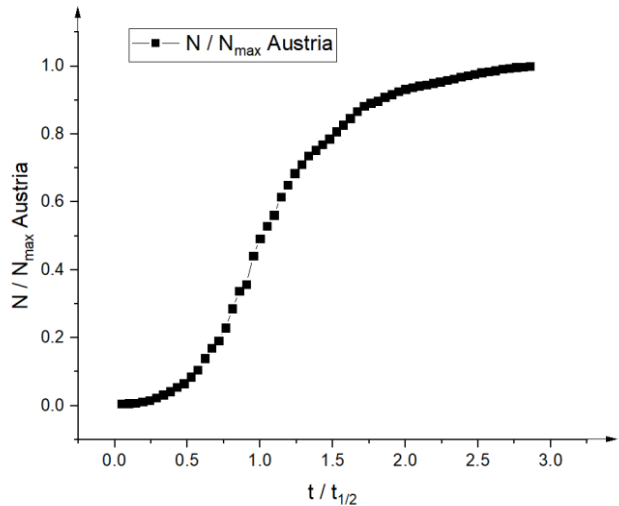 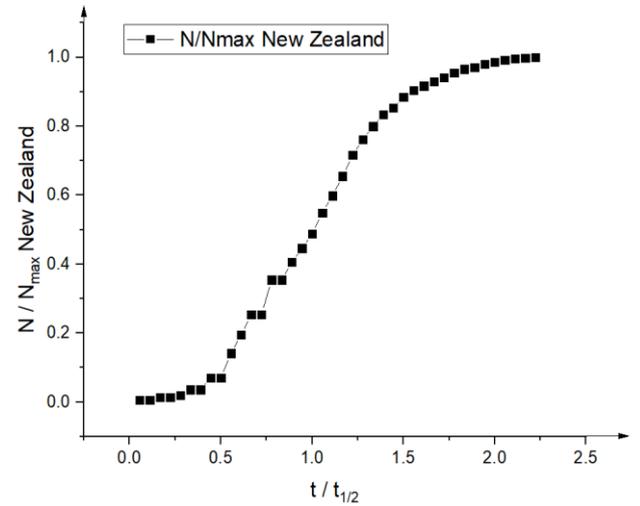

**Fig 2 a:** - We have N/Nmax of Austria in vertical axis and t / t$_{1/2}$ in horizontal axis.

**Fig 2 b:** - We have N/Nmax of New Zealand in vertical axis and t / t$_{1/2}$ in horizontal axis.

We focus on two specific countries ( one from Type-1 and the other from Type-2) in this work which we take as New Zealand and Austria respectively. We now re-plot the $N(t)$ vs $t$ data in a scaled form in Fig 2. We look at the value $N_{max}$ of $N(t)$ which is the value at the end of Phase -1. The ratio $r = N / N_{max}$ should not be sensitive to fluctuations in the testing efficiency etc and should be a somewhat robust quantity. We plot it along the y-axis. Along the x-axis we plot the number of "scaled" days. We do this scaling by taking the data from day zero to the day when $N = N_{max}$ and defining a half way point as $t_{1/2}$ the day where $N = N_{max}/2$. The variable $\tau = t/t_{1/2}$ is what we call "scaled "days and this is our x-axis. We will resort to this type of plotting in Sec IV.

In Fig 2 we show $r$ vs. $\tau$ for New Zealand and Austria and note that the primary difference is when the number of infections have become high ( $r > 0.5$). We attribute this fundamental difference between Type -1 and Type-2 largely to the ratio of asymptomatic carriers to symptomatic ones. In Sec.II we set up a Henon map to describe the propagation of the epidemic and in Sec.III discuss its fixed points and stabilities. We present the time series from the map in Sec.IV and give a brief conclusion. Some technical issues are discussed in Appendix A.

## 2. Basic model of Henon map:

On any given day we consider the population to consist of four distinct sets of people. The set {X} consists of people who can be infected. The set {Y} consists of people who have been infected but are asymptomatic while the set {Z} comprises the infected but symptomatic. The fourth set {Q} consists of people who are self-quarantined or hospitalized and are no longer part of the process. Instead of dealing with the actual number for each set we will consider the fractions obtained by dividing by the total population. We ignore the deaths which so far has fortunately been a very small fraction. We take three kinds of people at large on the n-th day– the healthy fraction, the asymptomatic infected fraction and the symptomatic infected fraction. The fourth fraction is not an active part of the dynamics. For practical purposes we identify {Y} as the set of people who are infected and hence transmitters but have not been tested as they have no symptoms. The set {Z} consists of those who have symptoms and have tested positive. We want to know how the fraction $y_n$ of infected but untested since asymptomatic people on day 'n' changes on day 'n+1'. The fraction $y_n$ will increase due to interaction of a member of set {X} with a member of set {Y} or {Z}. We assume that these interactions change {X} to {Y} with a probability $p$ and to {Z} with probability $1-p$. We also assume that the frequency of these interactions are proportional to $x_n y_n$ and $x_n z_n$ respectively. We assume that $y_n$ will decrease as the asymptomatic people get cured and also if there is an interaction between a member of set {Y} and a member of {Z} with a small probability of sending people to the {Z} class. This frequency we take to be proportional to $y_n z_n$. There will be some loss in $y_n$ due to recovery from infection and also may be due to contact tracing ( the contact tracing effect for no symptom people is negligible in reality ) as well and subsequent removal from the set {Y}. Thus, we get

$$y_{n+1} = y_n + p(ax_n y_n + bx_n z_n) - cy_n z_n - ly_n \qquad (1)$$

The number of people in the set {Z} increases with probability $1-p$ due to interactions between {X} and {Y} and between {X} and {Z}. It is possible that it can also increase due to interaction between {Y} and {Z} mentioned above. The loss in the number in this category is due to removal from scene on testing positive and recovery. We have

$$z_{n+1} = z_n + (1-p)[ax_n y_n + bx_n z_n] + cy_n z_n - fz_n \qquad (2)$$

It should be noted that the difference between $'l'$ and $'f'$ is a measure of the government's intervention. In general f would be somewhat higher because it is easier to quarantine the symptomatics but $'l'$ can become significant if the number of asymptomatics is higher and the government does a good job of tracing them quickly. Finally, we have the quarantined set {Q} who acquire people from the set {Z} and {Y} and through treatment or self-recovery ultimately restore (this effect we are not interested in because the set {Q} takes no active part in the dynamics) people to the healthy set X over a long-time scale. Consequently,

$$q_{n+1} = q_n + fz_n + ly_n - dq_n \qquad (3)$$

with the coefficient $d$ much smaller than the other coefficients. The change in $x_n$ occurs because of its interactions with {Y} and (Z)

$$x_{n+1} = x_n - (ax_n y_n + bx_n z_n) + dq_n \qquad (4)$$

ensuring the conservation law (since deaths have been ignored)

$$x_n + y_n + z_n + q_n = 1 \qquad (5)$$

At this point it is reasonable to assume that the $q_n$ is composed of some fraction of $y_n$ and some fraction of $z_n$ so that we can write using two coefficients $\alpha_1$ and $\alpha_2$

$$q_n = \alpha_2 z_n + \alpha_1 y_n \qquad (6)$$

This makes our conservation law constraint give

$$x_n = =1 - (1+\alpha_1)y_n - (1+\alpha_2)z_n \qquad (7)$$

We still have one more factor to contend with – the fact that the symptomatic patient may not be symptomatic for two to three days. This implies that the terms $x_n z_n$, $x_n y_n$ and also the term $y_n z_n$ must be augmented by terms such as $x_{n-1}z_{n-1}, x_{n-2}z_{n-2}$, $x_{n-1}y_{n-1}, x_{n-2}y_{n-2}, y_{n-1}z_{n-1}, y_{n-2}z_{n-2}$ etc. Instead of considering both subscripts $n-1$ and $n-2$, we consider only one term namely with subscript $n-1$ with an effective coefficient. We will refer to these terms with subscript $n-1$ as "delay" terms. The term $x_n(y_n + \eta z_n)$ will now be augmented by $\varepsilon x_{n-1}(y_{n-1} + \eta z_{n-1})$. We define b= $\eta a$, $c = a\beta$, $l = \mu a$ and $f = a\gamma$ so that "a" can be an adjustable scale parameter. Finally, we have our version of the Henon map as

$y_{n+1} = y_n + ap[y_n + \eta z_n][1 - (1 + \alpha_1)y_n - (1 + \alpha_2)z_n] + ap\varepsilon[1 - (1 + \alpha_1)y_{n-1} - (1 + \alpha_2)z_{n-1}][y_{n-1} + \eta z_{n-1}] - a\beta y_n z_n - a\beta\varepsilon y_{n-1}z_{n-1} - a\mu y_n$
(8)

$z_{n+1} = z_n + a(1-p)[y_n + \eta z_n][1 - (1 + \alpha_1)y_n - (1 + \alpha_2)z_n] + a(1-p)\varepsilon[1 - (1 + \alpha_1)y_{n-1} - (1 + \alpha_2)z_{n-1}][y_{n-1} + \eta z_{n-1}] + a\beta y_n z_n + \varepsilon a\beta y_{n-1}z_{n-1} - a\gamma z_n$
(9)

The parameter $\beta$ is expected to be much smaller than unity in the above equations. This factor comes from the possibility of an asymptomatic person becoming symptomatic on receiving more of the virus through an interaction with a symptomatic. The parameter $\varepsilon$ which describes the effect of delayed appearance of symptoms will have to be significantly smaller than unity. We define $\eta_1 = 1 + \alpha_1, \eta_2 = 1 + \alpha_2$ for subsequent use.

## III Fixed points and their stabilities

We begin by looking at the fixed points $(y^*, z^*)$ of the map written down in Eqs (8) and (9). These are the quantities of primary interest in the sense that $z^*$ denotes the total fraction of the symptomatic infected population in the full course of the epidemic which is obtained simply by testing symptomatic people for the virus. If P is the total population of the region considered ( city , state , country etc), then $z = N/P$ and the ratio $r$ introduced in the previous section is $r = z/z^*$. Similarly, the fraction of infected but asymptomatic people is $N_a/P$ where $N_a$ is the actual number people of this variety. The fixed point $y^*$ corresponds to the maximum value of the asymptomatic infected. The ratio $A = y^*/z^*$ is a very important parameter of the problem but not easily amenable to direct measurement.

We will see that $A$ can be inferred from the time series of the variable $z$. We begin by examining the fixed points of the map.

The fixed points ($y_0, z_0$) of the map given by Eqs (8) and (9) are found from the following relations

$$(1+\varepsilon)y_0(1-\eta_1 y_0) - (1+\varepsilon)[\eta\eta_1 + \eta_2 + \frac{\beta}{p}]y_0 z_0 - \frac{\mu}{p}y_0 + \eta(1+\varepsilon)z_0 - \eta\eta_2(1+\varepsilon)z_0^2 = 0 \quad (10)$$

And
$$(1+\varepsilon)y_0(1-\eta_1 y_0) + (1+\varepsilon)[\frac{\beta}{1-p} - \eta\eta_1 - \eta_2]y_0 z_0 + [\eta(1+\varepsilon) - \frac{\gamma}{1-p}]z_0 - \eta\eta_2(1+\varepsilon)z_0^2 = 0 \quad (11)$$

The stability of a fixed point ( $y_0, z_0$ ) can be found from the following set of equations obtained by linearizing the dynamics around the fixed point

$$\delta y_{n+1} = [1 + ap(1 - \eta_1 y_0 - \eta_2 z_0) - ap\eta_1(y_0 + \eta z_0) - a\beta z_0 - a\mu]\delta y_n + a[\eta p(1 - \eta_1 y_0 - \eta_2 z_0) - \eta_2 p(y_0 + \eta z_0) - \beta y_0]\delta z_n - \varepsilon ap[\eta_1(\eta z_0 + y_0) - (1 - \eta_1 y_0 - \eta_2 z_0) + (\beta/p)z_0]\delta y_{n-1} + \varepsilon a[p\eta(1 - \eta_1 y_0 - \eta_2 z_0) - p\eta_2(y_0 + \eta z_0) - \beta y_0]\delta z_{n-1} \quad (12)$$

$$\delta z_{n+1} = a[(1-p)\{1 - \eta_1 y_0 - \eta_2 z_0 - \eta_1(y_0 + \eta z_0)\} + \beta z_0]\delta y_n + \{1 + \eta a(1-p)[1 - \eta_1 y_0 - \eta_2 z_0] - a(1-p)\eta_2(y_0 + \eta z_0) + a\beta y_0 - a\gamma\}\delta z_n a\{\varepsilon(1-p)[1 - \eta_1 y_0 - \eta_2 z_0] + \beta\varepsilon z_0 - (1-p)\varepsilon(y_0 + \eta z_0)\eta_1\}\delta y_{n-1} + a\{\varepsilon(1-p)[1 - \eta_1 y_0 - \eta_2 z_0] + \beta\varepsilon y_0 - (1-p)\varepsilon(y_0 + \eta z_0)\eta_2\}\delta z_{n-1} \quad (13)$$

It is clear from Eqs (10) and (11) that there are four fixed points in general. One of them is easy to read off. It is the fixed point (0,0). The dynamics around it is obtained from Eqs (12) and (13) as

$$\delta y_{n+1} = (1 + ap - a\mu)\delta y_n + a\varepsilon p \delta y_{n-1} + ap\eta\delta z_n + \varepsilon\eta ap\delta z_{n-1} \quad (14)$$

and

$$\delta z_{n+1} = a(1-p)\delta y_n + [1 + \eta a(1-p) - a\gamma]\delta z_n + \varepsilon a(1-p)\delta z_{n-1} + \varepsilon a(1-p)\delta y_{n-1} \quad (15)$$

Trying the solutions $\delta y_n = A\lambda^n$ and $\delta z_n = B\lambda^n$, where A,B are constant coefficients and $\lambda$ is the growth rate or the 'eigenvalue', we get after elimination of A and B,

$$\{\lambda^2 - \lambda[1 + a(p - \mu)] - a\varepsilon p\}\{\lambda^2 - \lambda[1 - a\gamma + a\eta(1 - p)] - \varepsilon a(1 - p)\} - a^2 \eta p(1 - p)(\lambda + \varepsilon)^2 = 0 \quad (16)$$

The above is a quartic equation for the eigenvalue $\lambda$ whose magnitude has to exceed unity for the destabilization of the fixed point. In the absence of delayed response i.e. the small parameter $\varepsilon$ set exactly equal to zero, two roots are zero which says that for small $\varepsilon$, these two roots will be $O(\varepsilon)$ and hence play no role in the stability analysis. We obtain the other two roots at $\varepsilon = 0$ from the quadratic equation

$$\lambda^2 - \lambda[2 - a\gamma + a\eta(1 - p) + a(p - \mu)] + 1 - a\gamma + a\eta(1 - p)(1 - a\mu) + a(p - \mu)(1 - a\gamma) = 0 \quad (17)$$

For instability, we need the magnitude of $\lambda$ to be greater than unity and from the above quadratic equation for $\lambda$, this turns out to be

$$\gamma(p - \mu) + \mu\eta(1 - p) > 0 \quad (18)$$

The above condition is very important. Our primary focus in this work is the stability of the non-zero fixed point since we want to see how the number of infected people reaches a saturation value. But in the normal course of an epidemic it is clear that the end point is a return to the healthy state when all the infected people have been cured. The above condition tells us how that happens. Once we are very close to the non-trivial fixed point there will be extremely few people getting infected and consequently the recovery rates will be higher and also the probability of producing asymptomatics will go down and the above inequality will be violated allowing the origin to become a stable fixed point. The trivial (no infected people) and non-trivial (finite number of infected people) fixed points have finite basins of attraction in the parameter space and small changes in parameters ( which will happen when one is not encountering new cases of infection ) can take one from the vicinity of one to the other.

To understand the role of the delay terms which are the terms with the small parameter $\varepsilon$ in the coefficient, it suffices to consider the model with no asymptomatic patients. In other words, we set $y = 0$ and work only with the variable $z$. We now have (defining $a\eta(1 - p) = \alpha, a\gamma = \sigma, 1 + \alpha_2 = \nu$

$$z_{n+1} = z_n(1 - \sigma) + \alpha[z_n(1 - \nu z_n) + \varepsilon z_{n-1}(1 - \nu z_{n-1})] \quad (19)$$

For $\varepsilon = 0$ this is the standard logistic map. For $\varepsilon \neq 0$, we can call it a logistic map with delay. There are two fixed points of this map given by $x_1^* = 0$ (the trivial fixed point) and $x_2^* = [\alpha(1 + \varepsilon) - \sigma]/\alpha\nu(1 + \varepsilon)$ (the non-trivial fixed point). The trivial fixed point is unstable if $\alpha(1 + \varepsilon) > \sigma$. The nontrivial fixed point is the saturation fraction of infected people if there are no asymptomatics. The eigenvalue for this fixed point is $1 + \sigma - \alpha(1 + \varepsilon)$ which has to be less than unity for stability of the fixed point. This fixed point corresponding to a finite number of infected people will be approached more slowly as the parameters become such that the eigenvalue heads towards $-1$. A finite value of $\varepsilon$ ( by definition a positive quantity) takes the eigenvalue closer to $-1$ and thus delays the approach to saturation. Hence the physical role of the "delay" is to slow down the approach to the stabilization point which is the "end "of the infection.

We point out as an aside that the fixed point is destabilized if $\alpha(1+\varepsilon) > 2-\sigma$ and a two-cycle emerges. A change of variable casts Eq. (19) in the more recognizable form

$$u_{n+1} = \alpha[u_n(1-u_n) + \varepsilon u_{n-1}(1-u_{n-1})] \qquad (20)$$

This is for $\varepsilon = 0$, the well-studied Feigenbaum map where the trivial fixed point is destabilized at $\alpha = 1$ and the non-trivial fixed point is destabilized at $\alpha = 3$ giving rise to a two-cycle. The two cycle is de-stabilized at $\alpha = \sqrt{6}+1$ and a cascade of period-doubling bifurcations follow leading to an onset of chaos at $\varepsilon \cong 3.5699$. The same period doubling sequence is seen in Eq. (20) and is dealt with in Appendix A which concentrates on the role of the delay.

Having understood the role of the small parameter $\varepsilon$ in the dynamics, we can now simplify the dynamics of Eqs (8) and (9) by setting $\varepsilon = 0$. We also need to reduce the number of free parameters. The constant $\alpha_1$ which is a small appears in the combination $1+\alpha_1$ and then $\alpha_1$ can be dropped. The constant $\alpha_2$, on the other hand, is $O(1)$ and we will set it equal to unity. The parameter $\eta$ has been introduced to account for any difference between the nature of interaction between a healthy person and a symptomatic infected and a healthy person and an asymptomatic infected. As of now, this difference is not established and we will set $\eta = 1$. Our system consequently reduces to

$$y_{n+1} = y_n + ap(y_n + z_n)(1 - y_n - 2z_n) - a\beta y_n z_n - a\mu y_n \qquad (21)$$

$$z_{n+1} = z_n + a(1-p)(y_n + z_n)(1 - y_n - 2z_n) + a\beta y_n z_n - a\gamma z_n \qquad (22)$$

The parameter $a$ can be thought of as an overall scale factor that sets the time scale. The parameter $p$ is the probability of producing an asymptomatic infected person ($1-p$ is the probability of producing a symptomatic infected person), the parameter $\beta$ is the strength of the interaction between an asymptomatic and a symptomatic infected person with the interaction strength between a healthy and infected ( either variety ) set to unity. This parameter is expected to be small. The parameter $\gamma$ is the recovery rate combined with the quarantining on detection rate for the symptomatic infected while the parameter $\mu$ is a measure of the recovery rate of the asymptomatic infected. Having understood the qualitative effect of the delayed appearance of symptoms we have deliberately kept the small parameter $\varepsilon$ out the present discussion.

A constraint on the non-trivial fixed points is obtained from Eqs (21) and (22) as (for $z_0 \neq 0$)

$$z_0 = \frac{\mu y_0 (1-p)}{p\gamma - \beta y_0} \qquad (23)$$

If $\mu \to 0$, then $z_0 \neq 0$ implies $y_0 = \gamma p / \beta$ and if $\beta \to 0$, we end up with

$$z_0 = \frac{\mu y_0 (1-p)}{\gamma p} \qquad (24)$$

Use of Eq.(23) leads to the cubic equation given by

$$2\beta^2 z_0^3 - z_0^2[\beta^2(1-p\gamma) - 3\beta\gamma p - 4\mu\beta(1-p)] - z_0[\gamma p(\beta - p\gamma) + 2\mu\beta(1-p) - 3\mu\gamma p(1-p)$$
$$-2\mu^2(1-p)^2 - \mu\beta\gamma p(1-p) - \mu\beta\gamma] - \mu\gamma p(1-p) + \mu^2(1-p)(\gamma p - 1 + p) = 0 \quad (25)$$

The above cubic equation is not very transparent. The parameter $\beta$ is small but it quantifies an effect that has not been considered before. Its role can be examined best by studying the limit $\mu \to 0$. We will return to this speculative situation in a later publication. In the rest of this work we focus on $\beta = 0$.

This leads to the dynamical system

$$y_{n+1} = y_n + ap(y_n + z_n)(1 - y_n - 2z_n) - a\mu y_n \qquad (26a)$$

$$z_{n+1} = z_n + a(1-p)(y_n + z_n)(1 - y_n - 2z_n) - a\gamma z_n \qquad (26b)$$

The fixed points $(y_0, z_0)$ satisfy the constraint shown in Eq. (24) and now apart from the origin there is only one non-trivial fixed point which is given b

$$z_0 = \frac{\mu(1-p)}{p\gamma + 2\mu(1-p)} \times \frac{\mu(1-p) + \gamma(p-\mu)}{\mu(1-p) + p\gamma} \qquad (27)$$

To examine the stability of this fixed point, we need to linearize Eqs (26a) and (26b) about this fixed point and the eigenvalue $\lambda$ (instability if $|\lambda| > 1$) satisfies

$$\lambda^2 - Tr\lambda + Det = 0 \qquad (28)$$

with

$$Tr = 2 - a(\mu + \gamma) + a[p(1 - 2y_0 - 3z_0) + (1-p)(1 - 3y_0 - 4z_0)]$$

$$Det = (1 - a\gamma)^2 + a(1 - a\gamma)[p(1 - 2y_0 - 3z_0) + (1-p)(1 - 3y_0 - 4z_0)]$$

The instability condition is $Tr \geq 1 + Det$ and so long as the parameters are chosen such that the fixed point is stable, the evolution with the dynamics shown in Eqs (26a) and (26b) will give us the transmission dynamics of the type shown in Figs 2a and 2b. This is what we discuss in the next section.

## IV. Evolution of the Henon map trajectories and comparison with the trajectories of COVID-19 transmission

In this section we show the relevance of the Henon map shown in Eqs (26a) and (26b) by plotting the total fraction ($z_n / z_0$) of population who are infected and symptomatic (this is the number who test positive for COVID-19) against the number of iterations $n$ which corresponds to the number of days. The parameters have to be such that $z_n / z_0$ starts from a very small initial value (start of the epidemic) and then saturates at unity (reaches the fixed point of evolution). We have four parameters in our model of which $a$ is an overall scale factor for the duration of the evolution we will fix at unity for now (it will need to be adjusted when confronting real data). Thus, we need to vary three parameters $p, \gamma, \mu$ to get a feeling for how the evolution occurs. We present results for different parameter values in Figs (3-5).

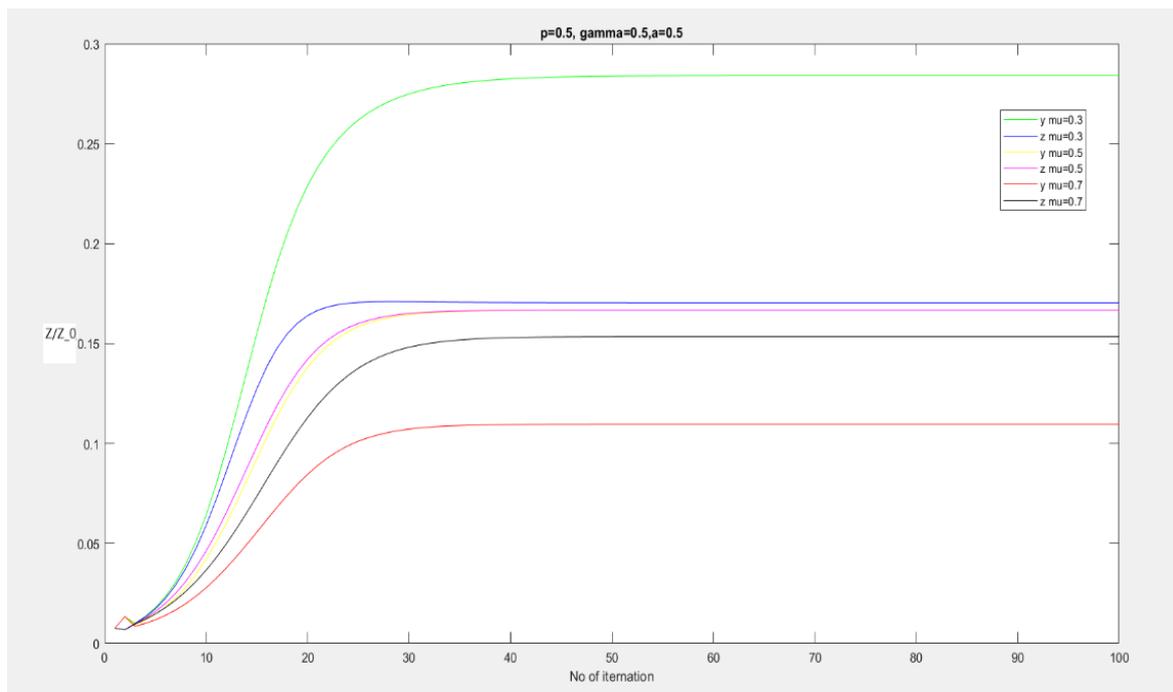

**Fig 3:** - $Z/Z_0$ Vs No of days with parameters $p=0.5, \gamma=0.5, a=0.5$

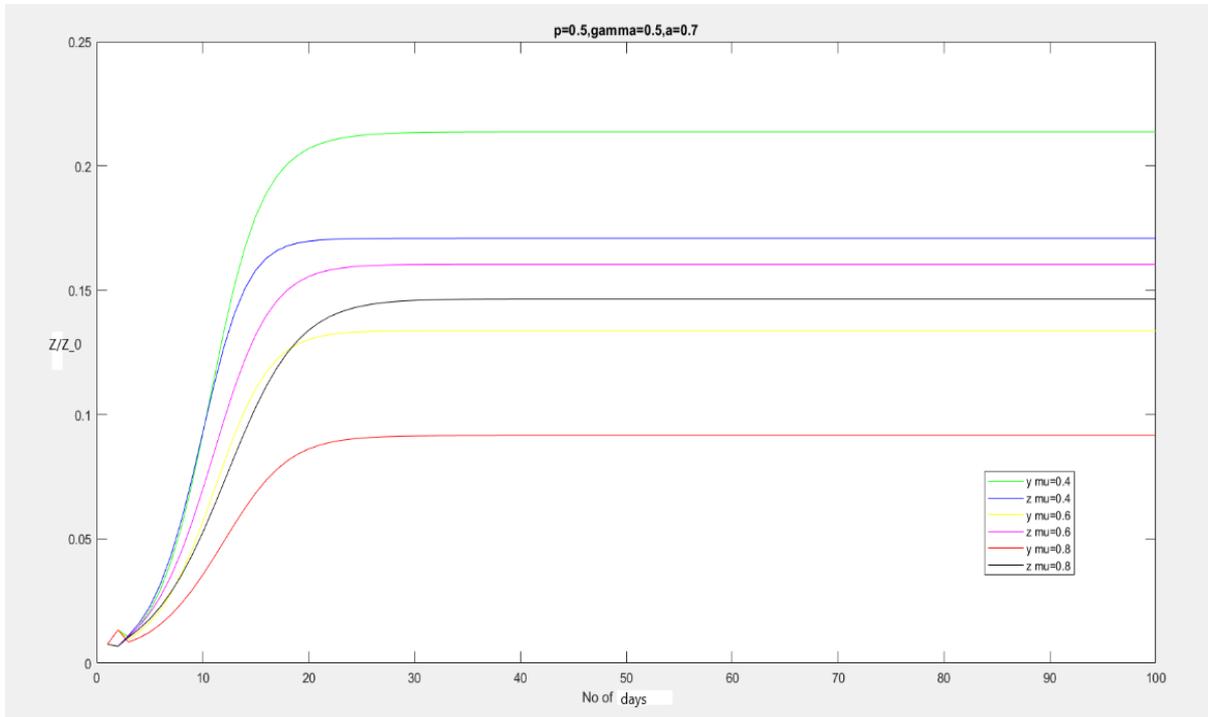

**Fig 4:** - Z/Z₀ Vs No of days with parameters
p=0.5, $\gamma = 0.5$, a $= 0.7$

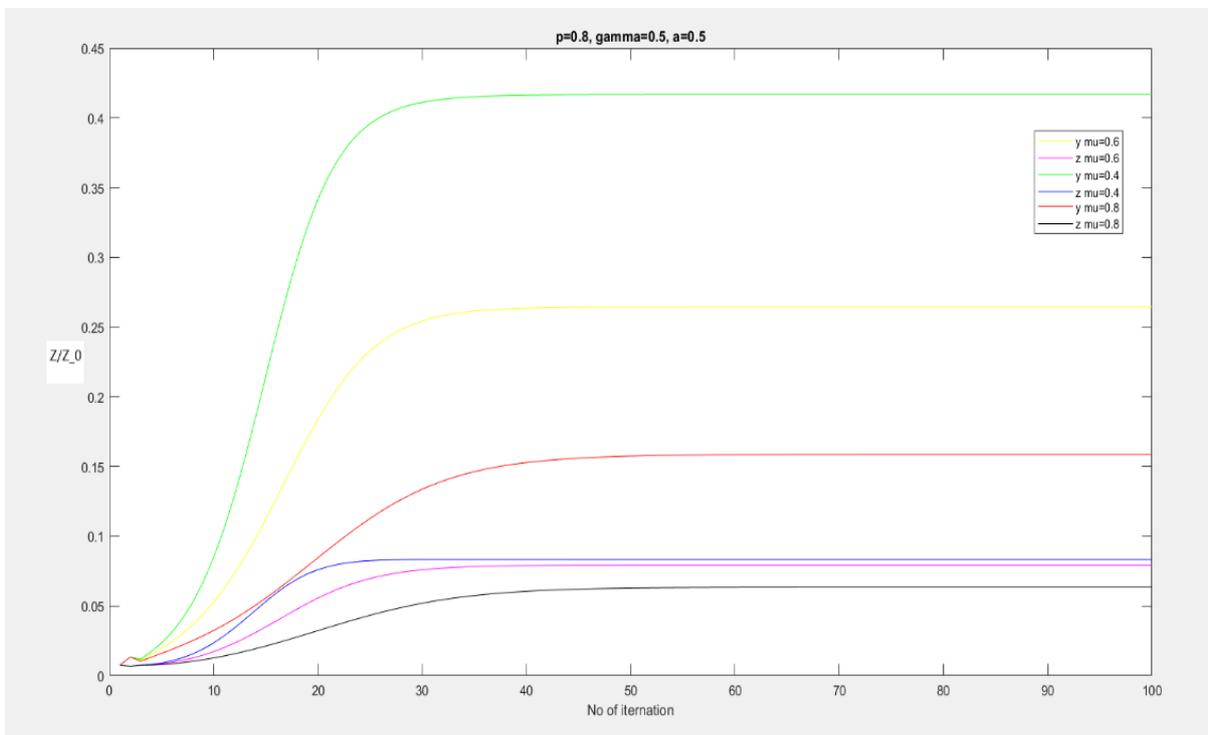

**Fig 5:** - Z/Z₀ Vs No of days with parameters
p=0.8, $\gamma = 0.5$, a $= 0.7$

The above evolution figures clearly indicate the following

From Figs (3) and (4), we see that with all other parameters remaining the same the parameter 'a' determines how spread out the evolution of the disease will be. This is natural as it was cast as an almost overall scale factor in Eqs (8) and (9). It is clear from Figs (3) and (5) that the parameter 'p' determines the initial growth rate. Larger the value of 'p', we expect from Eqs (8) and (9), that in the initial stages when the values of 'y' and 'z' are small the growth rate is proportional to (1-p) for the symptomatic patients. It is also apparent from Figs (3) and (5), that the saturation ratio of asymptomatic to symptomatic patients increases as the value of 'p' increases as seen clearly from Eq. (24). We also note that the removal rate 'gamma' of the symptomatic patients from the scene is vital in slowing down the spread of the disease. We see repeatedly from Figs 3-5 that if gamma is greater than mu, the saturation is reached faster for the symptomatic infected fraction. Thus, we establish the first two points mentioned in the abstract.

We now actually seek the parameter values which give a good description of the evolution of the pandemic in New Zealand and Austria. The results are shown in Figs 6 (Austria) and 7 (New Zealand). We have shown these plots against the scaled variable $\tau = t/t_{1/2}$ which we discussed in Sec I. This has the advantage of comparing the time evolutions of the two nations easily. It is clear from a study of Figs 6 and 7 that the pandemic saturates in New Zealand significantly faster than in Austria. It is also clear from these figures that for low values of $\varepsilon$ the course of evolution is not too different from the absence of this effect

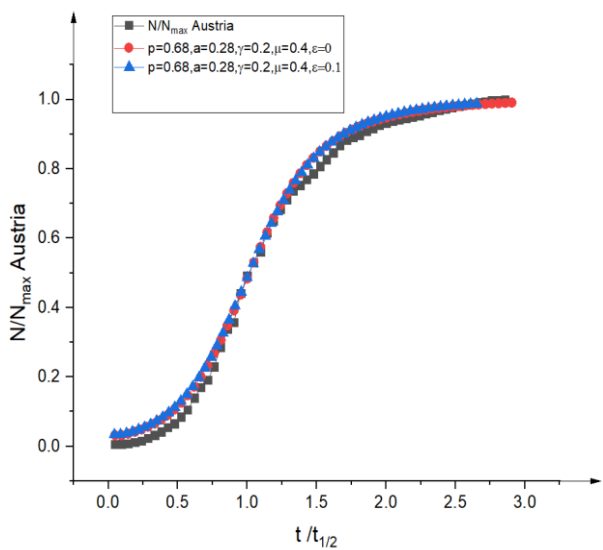
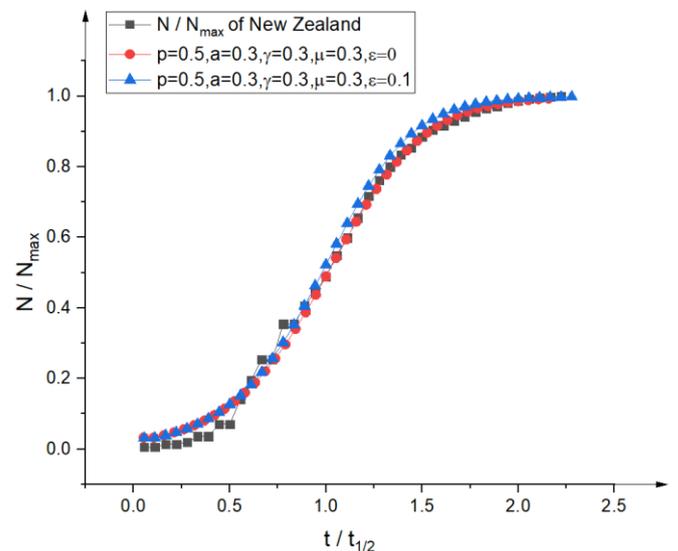

**Fig 6:** - The parameters for the prediction red curve are (p=0.6, a=0.42, $\mu = 0.6, \gamma = 0.5, \beta = 0, \varepsilon = 0.1$ and for blue curve (p=0.6, a=0.42, $\mu = 0.6, \gamma = 0.5, \beta = 0, \varepsilon = 0$)

**Fig 7:** - The parameters for the prediction blue curve are (p=0.5, a=0.3, $\mu = 0.5, \gamma = 0.5, \beta = 0, \varepsilon = 0$) and for red curve (p=0.5, a=0.3, $\mu = 0.5, \gamma = 0.5, \beta = 0, \varepsilon = 0.1$)

Given that we can find the evolution of the epidemic in Phase 1, the question arises –can we predict the course of the evolution for those countries which are yet to complete phase 1. This can be done by estimating the parameter "p" from the initial growth rate and since this corresponds to a ratio $p/(1-p)$ significantly greater than unity we use the curve for Austria as a universal curve to predict the future of the pandemic in India. We estimate that the $t_{1/2}$ corresponds to the last week of September 2020 and from that, using the logic of Sec I, conclude that the number would reach an end of phase-1 value of $(12.5 \pm 0.7) \times 10^6$ in the middle of March 2021. Thus, we arrive at the last point of our abstract. This should be compared with the result obtained by the SEIR model used by Agarwal et al [26]. We would like to point out that similar results have been obtained in our case with a far fewer number of parameters.

In conclusion we would like to point out that the technique used so far in modelling the Covid-19 dynamics have been SEIR models [26][27][28], delay differential equations [29] and the computer intensive agent-based model [30]. We present an alternative approach using maps which appears to be successful for the phase 1 of the disease spread.

## Appendix A: A logistic map with delay

We return to the Henon map as written down in Eqs. (8) and (9) and change it to a one-dimensional system by ignoring the presence of asymptomatics – this makes the system

$$z_{n+1} = z_n + a(1-p)\eta[z_n(1-\eta_2 z_n) + \varepsilon z_{n-1}(1-\eta_2 z_{n-1})] - a\gamma z_n \quad \text{(A1)}$$

The right-hand side has linear and quadratic terms in $z_n$ and $z_{n-1}$ we can carry out usual rescaling to write the system as

$$u_{n+1} = \rho[u_n(1-u_n) + c\varepsilon u_{n-1}(1-du_{n-1})] \quad \text{(A2)}$$

Without any loss of generality, we can set the numbers $c$ and $d$ equal to unity and work with the map (a different variation of a logistic map with delay)

$$u_{n+1} = \rho[u_n(1-u_n) + \varepsilon u_{n-1}(1-u_{n-1})] \quad \text{(A3)}$$

The fixed points of this map are $u_0 = 0$ and $u_0 = 1 - \dfrac{1}{\rho(1+\varepsilon)}$. Straightforward algebra shows that the former is destabilized for $\rho > (1+\varepsilon)^{-1}$. As for the latter, it is destabilized for $\rho > \rho_1 = (3-\varepsilon)/(1-\varepsilon^2)$, which can be shown by writing $u_n = u_0 + \delta u_n$ in Eq (A3) and linearising the resulting equation in $\delta u_n$. This yield

$$\delta u_{n+1} = \rho(1-2u_0)(\delta u_n + \varepsilon \delta u_{n-1}) \quad \text{(A4)}$$

Writing the increment $\delta u_n$ as $A\lambda^n$, we have

$$\lambda^2 = \lambda(\frac{2}{1+\varepsilon} - \rho) + \varepsilon(\frac{2}{1+\varepsilon} - \rho) \quad \text{(A5)}$$

Instability occurs when $\lambda = -1$ and gives the above value of $\rho_1$.

We now show that the state for $\rho > \rho_1$ is a periodic state of period 2 which can be represented as $u_n = A + (-)^n B$. Inserting this form in Eq. (A3) and equating terms with coefficient unity and terms with coefficient $(-)^n$ on either side of the resulting equation we obtain

$$A = \frac{1}{2}\left(1 + \frac{1}{\rho(1-\varepsilon)}\right) \quad \text{(A6)}$$

$$B^2 = \frac{A}{2\rho}\left(\rho - \frac{3-\varepsilon}{1-\varepsilon^2}\right) \quad \text{(A7)}$$

As expected Eq (A5) shows that the two cycle comes into existence at $\rho = \rho_1 = (3-\varepsilon)/(1-\varepsilon^2)$, the point where the fixed point loses stability. If we keep increasing the value of $\rho$, we anticipate that the two cycle loses stability and to obtain the critical value $\rho_2$ for this to happen, we set $u_n = A + (-)^n B + \delta u_n$ where $\delta u_n$ is an infinitesimal perturbation and ask for the linearized dynamics of $\delta u_n$. This leads to

$$\delta u_{n+1} = \rho[1 - 2\{A + (-)^n B\}]\delta u_n + \varepsilon\rho[1 - 2\{A + (-)^{n-1} B\}]\delta u_{n-1} \quad \text{(A8)}$$

We define $C = 1 - 2(A - B)$ and $D = 1 - 2(A + B)$. Anticipating that the 2-cycle will lose stability to a 4-cycle, we use $n = 2m+1, 2m, 2m-1$ and $2m-2$ obtain (to linear order)

$$\delta u_{2m+2} = \rho C \delta u_{2m+1} + \varepsilon \rho D \delta u_{2m} \quad \text{(A9a)}$$

$$\delta u_{2m+1} = \rho D \delta u_{2m} + \varepsilon \rho C u_{2m-1} \quad \text{(A9b)}$$

$$\delta u_{2m} = \rho C \delta u_{2m-1} + \varepsilon \rho D \delta u_{2m-2} \quad \text{(A9c)}$$

$$\delta u_{2m-1} = \rho D \delta u_{2m-2} + \varepsilon \rho C \delta u_{2m-3} \quad \text{(A9d)}$$

Our aim is to write a "period-doubled" form of Eq (A4) where we relate the $(2n+2)$ th increment to the $2n$ th and $(2n-2)$ th ones. We find from the above set

$$\delta u_{2n+2} = [\rho^2 CD + \varepsilon \rho D]\delta u_{2n} + \varepsilon \rho^3 C^2 D \delta u_{2n-2} \quad \text{(A10)}$$

This is the period doubled form of Eq (A4) and the growth rate $\lambda$ is given by

$$\lambda^2 = \left(\rho^2 CD + \varepsilon \rho D\right)\lambda + \varepsilon \rho^3 C^2 D \quad \text{(A11)}$$

Correct to $O(\varepsilon)$, the critical value $\rho_2$ for the onset of the 4-cycle ( obtained by setting $\lambda = -1$ ) is

$$\rho_2 = \sqrt{6} + 1 - \frac{\varepsilon}{\sqrt{6}} + O(\varepsilon^2) \qquad (A12)$$

The onset point for chaos through period doubling can be found by noting that two successive doublings follow from the conditions ( for $\varepsilon \ll 1$) from $2 - \varepsilon - \rho_1 = -1$ and $4 - 2\varepsilon - \rho_2^2 + 2\rho_2 = -1$. The merging of two successive bifurcations ( the onset of infinite period which is equivalent to onset of chaos ) would consequently occur (to $O(\varepsilon)$ ) for $\rho = \rho_c$ where

$$4 - \rho_c^2 + 2\rho_c - 2\varepsilon = 2(1-\varepsilon)^2 - \rho_c(1-\varepsilon) \qquad (A13)$$

To the lowest order in $\varepsilon$, the onset of chaos occurs at

$$\rho_c = \frac{3 + \sqrt{17}}{2} + \frac{\varepsilon}{2}(\frac{5}{\sqrt{17}} - 1) \qquad (A14)$$

We show below the 2-cycle at $\rho = 3.02$, the 4-cycle at $\rho = 3.48$ and chaos at $\rho = 3.68$ for $\varepsilon = 0.1$. It is interesting that while the onset of two cycle and four cycle is slightly delayed at $O(\varepsilon)$, the onset of period doubling chaos is slightly delayed beyond its $\varepsilon = 0$ value.

From Eq. A1 We get the following interesting curves:

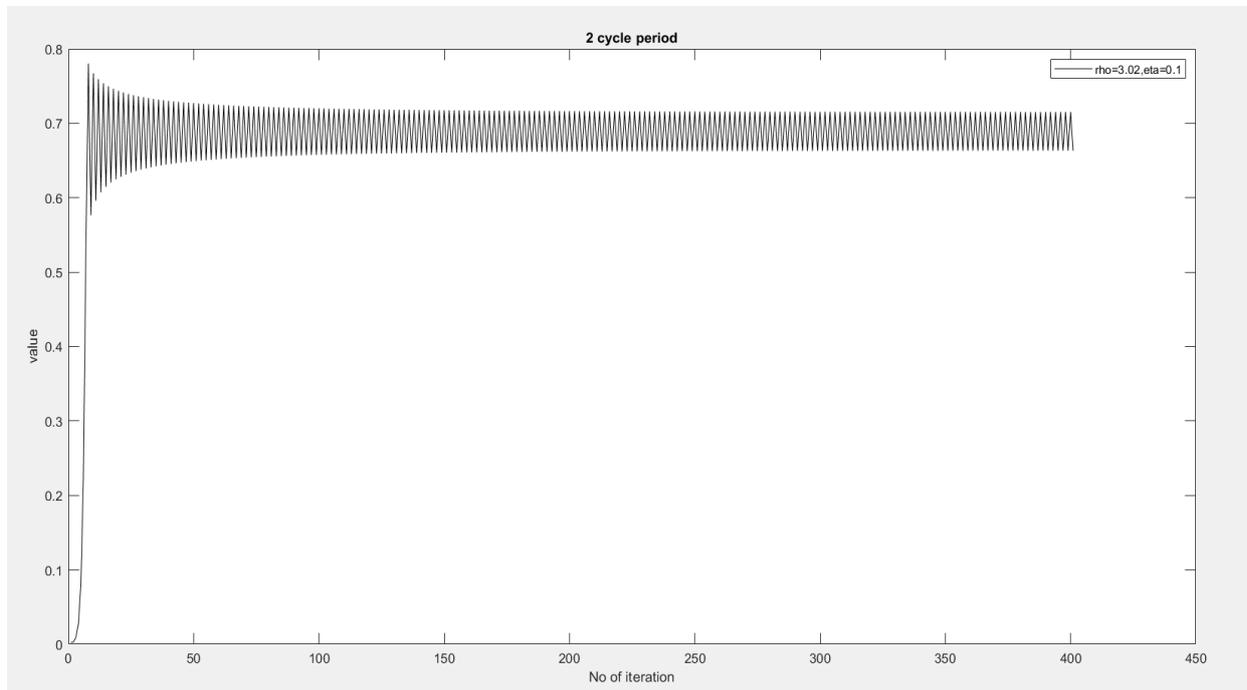

Fig 8: $\rho = 3.02$, $\varepsilon$=0.1 (2 cycle period)

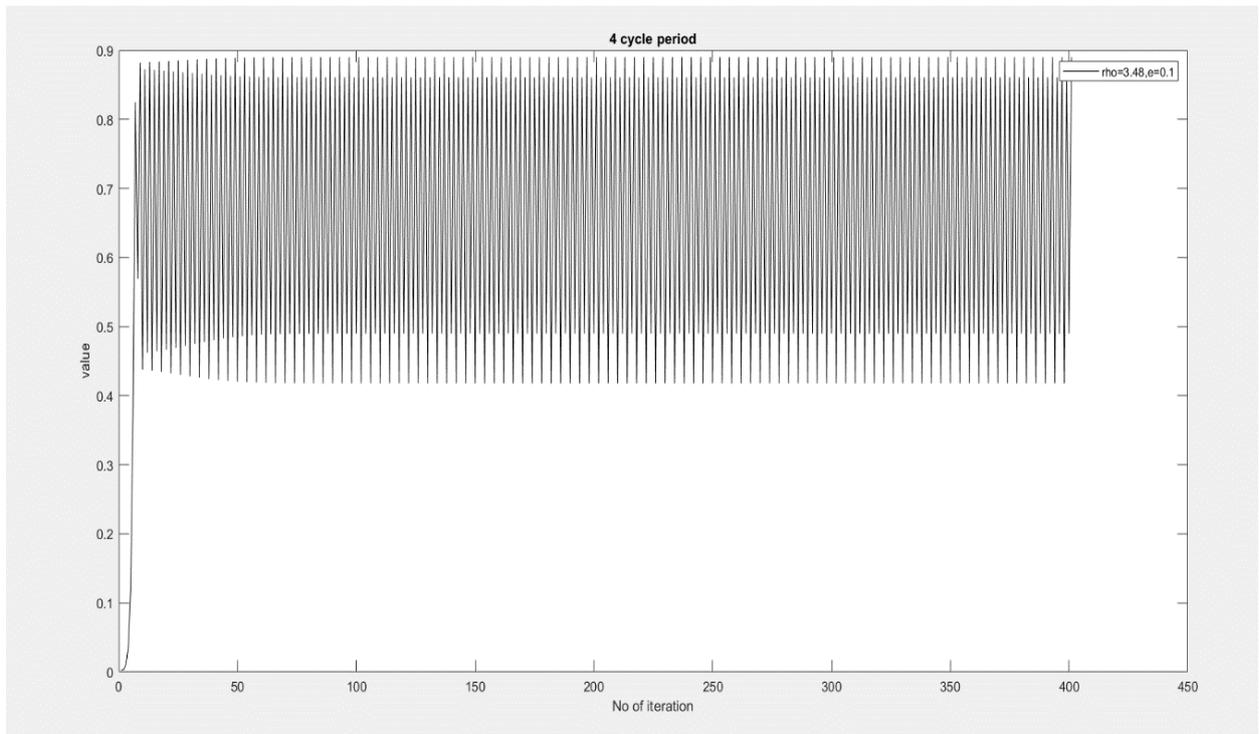

**Fig 9:** - $\rho = 3.48$, $\varepsilon$=0.1 (4 cycle period)

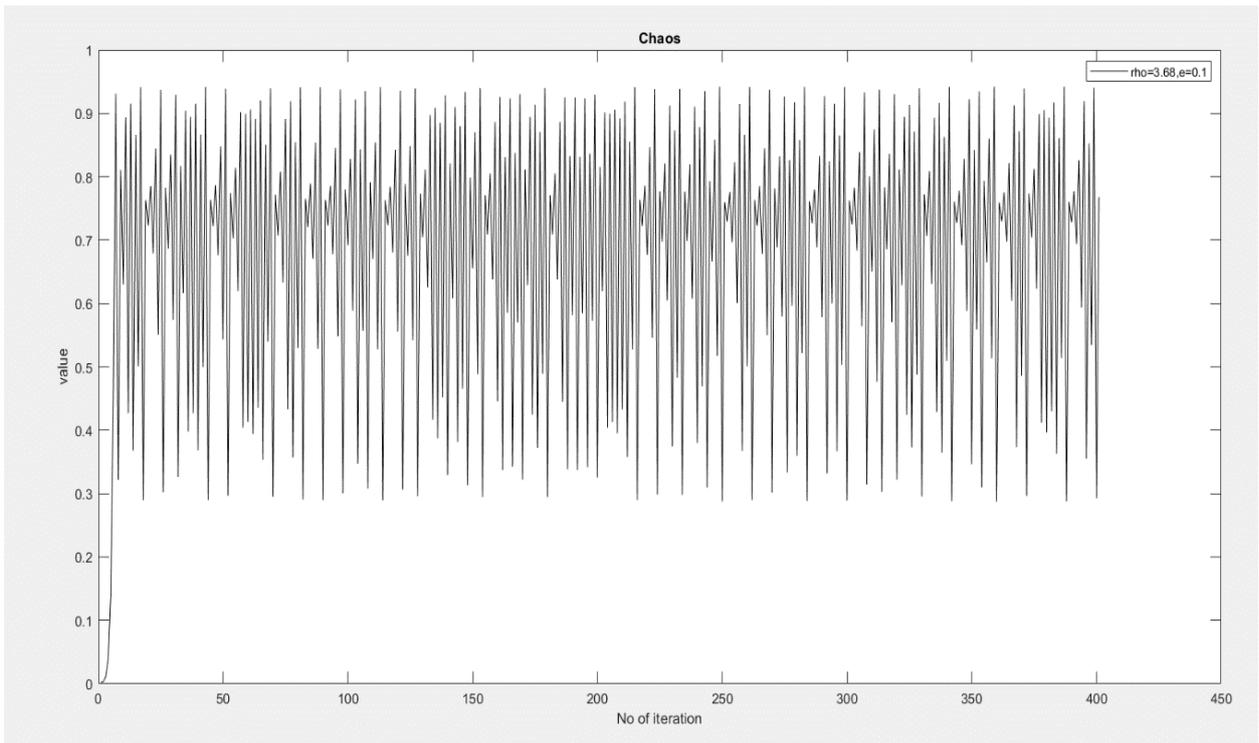

**Fig 10:** - $\rho = 3.68$, $\varepsilon$=0.1 (Chaos)

# REFERENCES

1. H. G. Schuster " Deterministic Chaos "  Physik –Verlag Publishers, Federal Republic of Germany ( 1984 )
2. R. Ross, " The prevention of malaria" published by John Murray ,  London (1911)
3. W.Kermach and A .G. McKendrick, Proc Roy. Soc. A115 , 700  (1927)
4. F. Brauer and C. Castillo-Chavez " Mathematical Models in Biology and Epidemiology" Springer
5. 5.H.W.Hethcote, " The Mathematics of Infectious Diseases" SIAM Review 42  No;4  599-653 (  2000  )
6. F.Brauer, Infectious Disease Modelling  2  113  (2017)
7. P. Chandra, IIT Kanpur Lecture Slides
8. J.L.Aron and I.B.Schwartz, J. Theo. Biol.  110   665 (1984)
9. W.M.Getz, R.Salter, O.Muellerklein, H.S.Yoon and K.Tallam,  Epidemics  25  9 (2018)
10. A.Godio,F.Pace and A.Vergnano, Int. J. Environ.Res.Pub.Health, 17   3535 (2020)
11. E.L. Piccolomini and F.Zama, PLOS  ONE  doi.org/10.1371/journal pone0237417 (2020)
12. S.Clemencon, V.C.Tran and H.D. Arazoza, Journal of Biological Dynamics 2:4 392-414 (2008)
13. L.J.S. Allen, Infectious Disease Modelling, 2  128 (2017)
14. A. J. Kucharski,T.W.Russell. C.Diamond,Y.Liu,J.Edmunds et al     The Lancet Infectious Diseases  20  553 (2020)
15. A.Simha,R.Venkatesha Prasad and S.Narayana, "A simple stochastic SIR model for COVID-19:Infection Dynamics for Karnataka after intervention-Learning from European trends"arXiv :2003.11920v3[q-bio.PE] April 16, 2020
16. S.He,Y.Peng and K.Sun, Nonlinear Dynamics doi.org/10.1007/s11071-020-05743-y (2020)
17. A.Das, A.Dhar, A.Kundu and S.Goyal  medRxiv 10.110/2020.06.04.20122580   (2020)
18. B.Shayak,  M.M.Sharma,R.H.Rand, A.K.Singh and A.Mishra " Transmission Dynamics of COVID-19 and Impact on Public Health Policy" 10.1101/2020.03.29.20047035 (2020)
19. B.Shayak and M.M.Sharma, doi 10.1101/2020.06.09.20126573 (2020)
20. B.Shayak and R.H.Rand, "Self- Burnout : A new path to the end of COVID-19 " doi 10.1101/2020.04.17.20069443 (2020)
21. I.Cherednik  arXiv : 2004.06021v3 April 20, 2020
22.  J. F. Rabajante, "Insights from early mathematical models of 2019.n-COV acute respiratory disease" arXiv article 2002.05296 (2020)
23.   M.Day, BMJ doi.org/10.1136/bmj.m1165 (2020)
24.  R.Anguelov, J.Banasiak, C.Bright, J.Lubuma and R.Ouifki , Biomath 9 2005103 (2020)
25. S.Jahedi and J. Yorke, medRxiv doi.og/10.1101/2020.06.23.20132522  (2020)
26. M.Agarwal,M.Kanitkar & M.Vidyasagar"Modelling the spread of SARS-Cov-2 pandemic impact of lockdown and interventions" Indian Journal of Medical Research.

27. G.Giordano et al " Modelling the COVID-19 epidemic and implementations of population-wide interventions in Italy"   Nature ( Medicine)  26 (6) 855-860  (2020)
28.  S.Raju " Did the Indian lockdown avert deaths ? "MedRxiv BMJ Yale https:/doi.org/10.1101/2020.06.27.20134932
29. B.Shayak,M.M.Sharma,R.H.Rand,A.Singh and A.Misra,"A Delay differential equation for the spread of COVID-19" International Journal of Engineering Research and Application 10 (1/3) 1-13 (2020)
30. S.Thurner, P.Klimer and R.Hanel, "A network based explanation of why most COVID-19 infection curves are linear" PNAS  117 (37) 22684-22689 (2020)